\documentclass[11pt]{article}
\pdfoutput=1

\usepackage{jheppub}

\usepackage{customprelude}

\title{Branes and Non-Invertible Symmetries}

\author{Iñaki García Etxebarria}

\emailAdd{inaki.garcia-etxebarria@durham.ac.uk}

\affiliation{Department of Mathematical Sciences, Durham University, Durham, DH1 3LE, United Kingdom}

\abstract{$\cN=4$ supersymmetric Yang-Mills theories with algebra
  $\fso(4N)$ and appropriate choices of global structure can have
  non-invertible symmetries. We identify the branes holographically
  dual to the non-invertible symmetries, and derive the fusion rules
  for the symmetries from the worldvolume dynamics on the branes.}

\begin{document}

\maketitle

\section{Introduction}

The notion of symmetry is undergoing rapid evolution: during the last
few years a number of works have convincingly argued that the classical
textbook definition of symmetry as a group of transformations acting
on local operators can (and should) be extended to include higher form
symmetries acting on extended operators \cite{Gaiotto:2014kfa}, higher
groups structures
\cite{Sharpe:2015mja,Tachikawa:2017gyf,Benini:2018reh} and more
generally higher categorical structures.

The importance of such higher categorical structures in two dimensions
has been realised for a long time, where they often appear from
discrete gauging
\cite{Frohlich:2009gb,Carqueville:2012dk,Brunner:2013xna,Bhardwaj:2017xup}.
A number of recent works have shown that symmetry operators without
inverses (and which are therefore not elements of any group, but
should rather be thought of in categorical terms) are also very common
in higher dimensional theories
\cite{Gaiotto:2019xmp,Heidenreich:2021xpr,Choi:2021kmx,Kaidi:2021xfk,
  Roumpedakis:2022aik,
  Bhardwaj:2022yxj,Arias-Tamargo:2022nlf,Choi:2022zal,Choi:2022jqy,
  Cordova:2022ieu,Kaidi:2022uux,Antinucci:2022eat,Bashmakov:2022jtl,Damia:2022bcd,Bhardwaj:2022lsg,Lin:2022xod,Bartsch:2022mpm}. In
this paper we will focus on one class of theories where such
non-invertible symmetries appear: $\cN=4$ theories with gauge
group\footnote{In this note we do not aim to analyse fully the mapping
  from boundary conditions to global structures, so we will ignore the
  existence of discrete choices of $\theta$ angles in some of the
  theories we discuss \cite{Aharony:2013hda}. A careful analysis of
  the mapping from global structures to properties of the holographic
  duals will be provided in \cite{EGEHR}.} $\Pin^+(4N)$, $Sc(4N)$ and
$PO(4N)$ \cite{Bhardwaj:2022yxj}. The details are a little different
in the three cases, so in this introduction we will focus on the
$Sc(4N)$ case for concreteness. This theory has three 2-surface
symmetry generators, which we will call $D_2^{\ssc,e}(\Sigma_2)$,
$D_2^{\sss,m}(\Sigma_2)$ and their product
$D_2^{\ssc,e}(\Sigma_2)D_2^{\sss,m}(\Sigma_2)$. There is additionally a
three-surface operator $\cN(\cM^3)$. The terms in the fusion algebra
involving $\cN(\cM^3)$ are
\begin{subequations}
  \label{eq:Sc-fusion-intro}
  \begin{align}
    \label{eq:Sc-fusion-intro-N^2}
    \cN(\cM^3)\times \cN(\cM^3) & = \sum_{\Sigma_2,\Sigma_2'\in H_2(\cM^3;\bZ)}
                                  D_2^{\ssc,e}(\Sigma_2) D_2^{\sss,m}(\Sigma_2')\, ,\\
    \cN(\cM^3)\times D_2^{\ssc,e}(\Sigma_2) & = \cN(\cM^3)\, ,\\
    \cN(\cM^3)\times D_2^{\sss,m}(\Sigma_2) &= \cN(\cM^3)\, .
  \end{align}
\end{subequations}
The right hand side of~\eqref{eq:Sc-fusion-intro-N^2} is generically a
sum of operators, and therefore $\cN(\cM^3)$ is not invertible.

All these theories can be obtained from the $\cN=4$ $SO(4N)$ theories
by suitable gaugings of discrete symmetries. Whether we have performed
the gauging or not is not visible for a local observer measuring
processes on a topologically trivial (but arbitrarily large)
neighbourhood of a point. This suggests that the holographic dual of
all these theories is the same, which is indeed the case: the
holographic dual is in all cases IIB on $\AdS_5\times \RP^5$. The
different theories arise from different choices of asymptotic
behaviour for discrete gauge fields in the bulk, as discussed in
related examples in \cite{Witten:1998xy,Aharony:2016kai}.

Since all these theories share the same bulk description, it should be
possible to describe the non-invertible symmetry generators (in the
cases where they are present in the field theory) in terms of objects
living on the holographic IIB dual. The goal of this note is to
identify these objects, and to derive their fusion rules using IIB
techniques.\footnote{The techniques we use in our analysis do not
  require knowledge of the Lagrangian of the boundary SCFT (although
  the choice of theories to study is certainly informed by the field
  theory results in \cite{Bhardwaj:2022yxj}, and we will chose our
  notation to dovetail the field theory analysis), so they apply
  equally well to the study of non-Lagrangian theories realised either
  holographically or via geometric engineering. See
  \cite{Bashmakov:2022jtl} for a recent study of non-invertible
  symmetries in non-Lagrangian theories using a different
  approach.}$^{,}$\footnote{We refer the reader to
  \cite{Damia:2022bcd} for a holographic study of a different class of
  non-invertible defects.}  Surprisingly, given the perhaps unfamiliar
fusion relations~\eqref{eq:Sc-fusion-intro}, it will transpire that
the symmetry generators are represented holographically by ordinary
branes wrapping torsional cycles in the internal $\RP^5$.

In order to explain how this is possible, it is useful to review
briefly how the fusion relations~\eqref{eq:Sc-fusion-intro} are
derived in \cite{Bhardwaj:2022yxj} (see also \cite{Kaidi:2021xfk}). We
start with the $SO(4N)$ theory, which has a $\bZ_2$ outer automorphism
0-form symmetry and a $\bZ_2\times\bZ_2$ 1-form symmetry. We will
denote the background for the 0-form symmetry $A_1$, and the
backgrounds for the two $\bZ_2$ factors $B_2^m$ and $C_2^e$. There is
a cubic 't Hooft anomaly represented by an anomaly theory with action
\[
  \label{Anomaly}
  i\pi \int A_1 B_2^{e} C_2^{m} \,.
\]
We obtain the $Sc(4N)$ theory by gauging both 1-form symmetries
simultaneously.\footnote{There is a discrete choice when gauging,
  related to the precise way in which we sum over $B_2^m$
  backgrounds. A slightly different choice (related by the outer
  automorphism) gives the $Ss(4N)$ global form instead, which also has
  non-invertible symmetries. The analysis of both cases is essentially
  identical, so we will focus on the $Sc(4N)$ case.} (The $\Pin^+$ and
$PO$ cases are obtained by gauging other pairs of symmetries involved
in the cubic anomaly.)  Naively, we would say that the 0-form symmetry
is broken due to the cubic anomaly~\eqref{Anomaly}. The more precise
statement is that due to the anomaly the generator $D_3^{(0)}(\cM^3)$
of the 0-form symmetry is not invariant under combined gauge
transformations of $B_2^m$ and $C_2^e$. But as argued in
\cite{Kaidi:2021xfk,Bhardwaj:2022yxj} it is possible to ``dress'' (or
stack) $D_3^{(0)}(\cM^3)$ with an anomalous TQFT $\cT$ depending on
$B_2^m$ and $C_2^e$. The combined topological operator is gauge
invariant, and survives as a topological operator of the gauged
theory. The price to pay is that the fusion rules for $\cT$ are more
involved, and lead to non-invertibility of the dressed operator
$D_3^{(0)}(\cM^3)\times \cT$ (the details will be reviewed below).

Coming back to the IIB holographic setup, the main observation of this
paper is that $D_3^{(0)}(\cM^3)\times\cT$ is precisely the IR limit of
the theory on branes wrapping suitable torsional cycles in the
holographic dual background. For instance, we will see that in the
$Sc(4N)$ case the non-invertible operator arises from a D3 brane
wrapping $\cM^3\times \RP^1$, where $\RP^1\subset \RP^5$. Reducing on
the $\RP^5$ leaves an effective 3-dimensional brane wrapping $\cM^3$
inside the five dimensional bulk, which becomes $D_3^{(0)}(\cM^3)$
when pushed to the boundary.

A pleasing consequence of the identification in this note is that
anomaly cancellation of the dressed operator follows automatically:
the background fields for the symmetries of the theory are given by
asymptotic values for the supergravity fields in the IIB dual, and the
D3 brane action is necessarily gauge invariant under all possible
gauge transformations of these (although the precise way in which this
happens is often subtle). Since anomaly cancellation is automatic once
we start talking about branes, it is illuminating to understand why
non-invertible symmetries appear in the holographic dual without
referring to anomalous operators. This is also desirable since the
split between the bare $D(\cM^3)$ and its ``dressing'' $\cT$ is
unnatural in terms of the brane theory, particularly once we try to
formulate things in the language of K-theory. We provide such an
explanation below in terms of incomplete cancellation of induced brane
charges due to quantum effects.

\subsubsection*{Note added}

\emph{I thank the authors of \cite{ABBS} for informing me of their related
upcoming work, where they give complementary evidence for the
identification of non-invertible symmetries with branes in holographic
settings, and for agreeing to coordinate submissions.}

\section{4d $\cN=4$ $\mathfrak{spin}(4N)$ SYM and non-invertibles}

The $\Spin(4N)$ SYM theory has a 2-group structure, with one-form
symmetry group\footnote{Our conventions are as follows: $\Spin(4n)$
  has two spinor irreps unrelated by complex conjugation, which we
  denote by ``$\sss$'' and ``$\ssc$''. $\bZ_2^\sss$ acts on $\ssc$,
  and leaves $\sss$ invariant, while $\bZ_2^\ssc$ acts on $\sss$ and
  leaves $\ssc$ invariant. This choice of notation is motivated by
  consistency with the fact that the diagonal $\bZ_2$ combination,
  traditionally denoted $\bZ_2^V$, does not act on the vector. We
  define $Sc(4N)\df \Spin(4N)/\bZ_2^\sss$. \label{fn:conventions}}
\begin{equation}
\Gamma^{(1)} = \mathbb{Z}_2^{\sss} \times \mathbb{Z}_2^\ssc \,,
\end{equation}
and a 0-form symmetry part $\mathbb{Z}_2^{(0)}$ which is an outer
automorphism that acts on the 1-form symmetry by exchanging the two
factors: $\mathbb{Z}_2^\sss \leftrightarrow \mathbb{Z}_2^\ssc$.  We
will now construct the topological defects that generate these
symmetries in the holographic dual.

This holographic dual is obtained as the near horizon limit of a stack
of D3-branes on top of an O3$^-$ orientifold, and is given by IIB
string theory on $\AdS_5\times \RP^5$ \cite{Witten:1998xy}. In general
we want to put the field theory on some spin\footnote{We will assume
  for simplicity that neither $\cM^4$ nor any of the submanifolds
  where we will wrap defects contains torsion in homology. This is not
  physically required, but it simplifies some of the formulas below.}
manifold $\cM^4$ different from $S^4$, so we will replace $\AdS_5$ by
a non-compact manifold $X^5$ which asymptotically becomes
$\bR\times \cM^4$ \cite{Witten:1998wy}. There is a non-trivial
$\SL(2,\bZ)$ duality fibration over $\RP^5$, which acts with the
$-1\in \SL(2,\bZ)$ element as we go around the non-trivial generator
of $\pi_1(\RP^5)=\bZ_2$. (This element can be represented
alternatively as $\Omega F_L$ in worldsheet terms, but with future
generalisations in mind we will describe it as an $SL(2,\bZ)$ bundle
instead.)  The 2-form supergravity fields $B_2$ and $C_2$ get a sign
under this action, and project down to $\bZ_2$ fields on $\AdS_5$,
while $C_4$ does not get a sign and survives as a continuous field. We
will find it useful to work in a democratic formulation, where we also
include the $B_6$ and $C_6$ fields magnetic dual to $B_2$ and
$C_2$. $\SL(2,\bZ)$ is a gauge symmetry of the theory on the
(orientable) space $\AdS_5\times \RP^5$, so in order for the action to
be well defined we need $B_6$ and $C_6$ to also transform with a minus
sign under $-1\in \SL(2,\bZ)$.

What this means is that $H_3$ and $F_3$ are elements of the cohomology
group with local coefficients $H^3(X^5\times \RP^5; \tbZ)$ (we refer
the reader to appendix 3.H of \cite{Hatcher:478079} for details), and
similarly their magnetic duals $H_7$ and $F_7$ are elements of
$H^7(X^5\times\RP^5;\tbZ)$. On the other hand $F_5$ is classified by
$H^5(X^5\times \RP^5;\bZ)$. In what follows we will focus on the
structure on $\RP^5$, as the $SL(2,\bZ)$ bundle is trivial on
$X^5$. The untwisted cohomology groups of $\RP^5$ are standard, and
the twisted ones can be derived easily from the results in
\cite{Thom1952}:
\begin{equation}\label{RP5Coho}
\begin{aligned}
H^\ast (\RP^5, \mathbb{Z}) & = \{\mathbb{Z}\,,\ 0\,,\  \mathbb{Z}_2 \,,\ 0 \,,\ \mathbb{Z}_2 \,,\ \mathbb{Z} \}\cr 
H^\ast (\RP^5, \widetilde{\mathbb{Z}}) & = \{ 0\,,\  \mathbb{Z}_2\,,\ 0 \,,\ \mathbb{Z}_2 \,,\ 0\,, \ \mathbb{Z}_2 \}\,.
\end{aligned}
\end{equation}

Similar considerations hold for homology: $(p,q)$ 1-branes (such as
fundamental strings and D1 branes) are elements of
$H_2(X^5\times\RP^5;\tbZ)$, $(p,q)$ 5-branes are elements of
$H_6(X^5\times\RP^5;\tbZ)$, and D3 branes are elements of
$H_4(X^5\times\RP^5;\bZ)$. The relevant homology groups are (by
Poincaré duality, which holds since $\RP^5$ is orientable)
\begin{equation}\label{RP5homology}
\begin{aligned}
H_\ast (\RP^5, \mathbb{Z}) & = \{\mathbb{Z}\,,\ \bZ_2\,,\  0 \,,\ \bZ_2 \,,\ 0 \,,\ \mathbb{Z} \}\cr 
H_\ast (\RP^5, \widetilde{\mathbb{Z}}) & = \{ \bZ_2\,,\  0\,,\ \bZ_2 \,,\ 0 \,,\ \bZ_2\,, \ 0 \}\,.
\end{aligned}
\end{equation}

With an understanding of the cycles that the branes can wrap, it is
straightforward to identify the charged operators of the $\Spin(4N)$
theory \cite{Witten:1998xy}: the vector Wilson line $W_V$ is a
fundamental string on a point of $\RP^5$, the $\sss$-spinor Wilson
line $W_\sss$ is a D5-brane on $\RP^4$, and finally the $\ssc$-spinor
Wilson line $W_\ssc$ is the combination of both previous lines: a
D5-brane/F1 bound state, again wrapped on $\RP^4$. In all cases the
branes wrap a surface on $X^5$ extending to the boundary, where they
end on a line.

We can go to the $SO(4N)$ theory by gauging the diagonal factor
$\bZ_2^V \subset \bZ_2^\sss\times \bZ_2^\ssc$. The vector line $W_V$
is unaffected by the gauging, so it survives, but the $W_\sss$ and
$W_\ssc$ lines are no longer gauge invariant, and become non-genuine
(that is, boundaries of surface operators). A non-genuine line $H_V$
of the $\Spin(4N)$ theory, with $w_2^\sss=w_2^\ssc$ flux around it,
now becomes a genuine line operator in the $SO(4N)$
theory. Holographically this operator corresponds to a D1 brane
wrapping a point in $H_0(\RP^5;\tbZ)=\bZ_2$.\footnote{The simplest
  derivation of this fact follows from recalling that the $SO(4N)$
  field theory is invariant under $SL(2,\bZ)$, which maps to an
  $SL(2,\bZ)$ action on the holographic dual. We refer the reader to
  \cite{EGEHR} for a systematic analysis.} We denote the generators
for these two symmetries $D_2^{B,e}(\cM^2)$ (acting on fundamental
strings) and $D_2^{C,m}(\cM)$ (acting on D1 branes), and the
corresponding background fields $B_2^e$ and $C_2^m$.

Starting from the $SO(4N)$ theory we can gauge various pairs of global
symmetries, an operation that, due to the cubic anomaly~\eqref{Anomaly},
results in theories with non-invertible symmetries
\cite{Kaidi:2021xfk,Bhardwaj:2022yxj}:
\begin{equation}\label{GlobalForms}
\begin{aligned}
  \Pin^+(4N)\colon &\quad \text{gauge $D_3^{(0)}$ and $D_2^{C,m}$} \\
  Sc(4N)\colon & \quad \text{gauge $D_2^{B,e}$ and $D_2^{C,m}$} \\
  PO(4N)\colon & \quad \text{gauge $D_3^{(0)}$ and $D_2^{B,e}$} \,.
\end{aligned}
\end{equation}
In these expressions $D_3^{(0)}$, or more precisely $D_3^{(0)}(\cM^3)$
is the generator for the outer automorphism 0-form symmetry of the
$SO(4N)$ theory.

We are thus led to the crucial question in this paper: having
identified the \emph{charged} operators in the field theory in terms
of the holographic dual, what is the holographic description of the
\emph{charge} operators implementing the global symmetries in the
$SO(4N)$ theory?

For concreteness, let us specialise to the holographic dual of the
symmetry generator $D_2^{C,m}(\cM^2)$ of the $SO(4N)$ theory,
measuring how many 't Hooft lines (mod 2) $H_V$ are linked by $\cM^2$,
without taking into account the Wilson lines $W_V$. Given our
identification of lines above, a natural guess would be
\begin{equation}
  \label{eq:D2V-try1}
  D_2^{C,m} (\cM^2) \stackrel{?}{\to} e^{i\pi \int_{\cM_2\times \RP^4} C_6}\, ,
\end{equation}
where $\cM^2$ lives on $\cM^4$, and becomes the symmetry operator when
pushed to the boundary. This holonomy certainly measures the number of
D1 branes linked by $\cM^2$ (the basic argument is given below in case
of the outer automorphism 0-form symmetry), but it cannot be the right
answer for a number of reasons. First, we know that in IIB string
theory fluxes are not measured by cohomology, but rather K-theory
\cite{Moore:1999gb,Freed:2000tt,Freed:2000ta}. A way of capturing the
right K-theoretic formula is to phrase the answer in terms of the
Wess-Zumino coupling in the D5 brane action:
\begin{equation}
  \label{eq:D2V-try2}
  D_2^{C,m} (\cM^2) \stackrel{?}{\to} e^{\text{WZ}(\cM_2\times \RP^4)}\, .
\end{equation}
where \cite{Cheung:1997az,Minasian:1997mm,Freed:2000ta}
\begin{equation}
  \label{eq:WZ}
  \mathrm{WZ}(X) = 2\pi i \int_X e^{F_2-B_2} \sqrt{\frac{\hat{A}(TX)}{\hat{A}(NX)}} (C_0 + C_2+\ldots)
\end{equation}
A second reason why we expect neither \eqref{eq:D2V-try1} nor
\eqref{eq:D2V-try2} to be the full answer is that in string theory
there are no local operators, only dynamical objects. So we should aim
to represent the charge generator by a dynamical object, and not
simply a defect. The dynamical objects that are electrically charged
under $C_6$, and would arise when fixing the insertion of the
defect as a boundary condition, are D5 branes.

While neither argument is conclusive, they both suggest that the
holographic description of the symmetry generator is a full D5, pushed
to the boundary:\footnote{The authors of \cite{ABBS} provide
  complementary evidence for the same proposal.}
\begin{equation}
  \label{eq:brane-ansatz}
  D_2^{C,m} (\cM^2) \to \text{D5}(\cM_2\times \RP^4)\, .
\end{equation}
This ansatz has the additional virtue of restoring the common
origin between lines and charge generators, familiar from the
formulation of symmetries in terms of relative field theories
\cite{Freed:2012bs}.

An objection one might raise about~\eqref{eq:brane-ansatz} is that
branes are not topological, while charge operators should be. As we
will see in a moment, the worldvolume theory on the branes, when
reduced to $\cM^2\subset X^5$, is a discrete $\bZ_2$ gauge
theory. Therefore the potential lack of deformation-invariance coming
from the gauge fields on the brane is not an issue. There is still an
overall factor of the volume, but it does not couple to the dynamical
fields of the field theory on the boundary, so it can be absorbed into
a counterterm.

A subtle feature of~\eqref{eq:brane-ansatz} is that the worldvolume
theories on the brane are quantum field theories, so we should sum
over them. As we will argue, the sum over worldvolume degrees of
freedom provides precisely the minimal anomalous TQFT ``dressing'' the
bare symmetry generator identified in \cite{Bhardwaj:2022yxj}. This is
a very non-trivial test of the identification~\eqref{eq:brane-ansatz}.

Clearly, if the ansatz~\eqref{eq:brane-ansatz} is correct, the
holographic dual of the operator counting 't Hooft lines $H_V$ is the
S-dual of~\eqref{eq:brane-ansatz}:
\begin{equation}
  \label{eq:NS5-brane-ansatz}
  D_2^{B,e}(\cM^2) \to \text{NS5}(\cM_2\times \RP^4)\, .
\end{equation}
Additionally, the $SO(4M)$ theory has the 0-form parity symmetry
discussed above. The point operator charged under this symmetry is
known as the Pfaffian operator. As discussed in \cite{Witten:1998xy}
the Pfaffian operator is represented holographically by a D3 brane
wrapping the $\RP^3$ cycle inside $\RP^5$, and extending to a point on
the boundary. We will refer to this brane as the ``Pfaffian brane''.

We now argue that the holographic dual of the generator of this
symmetry is
\begin{equation}
  \label{eq:D3-generator}
  D_3^{(0)}(\cM^3) \to \text{D3}(\cM^3\times \RP^1)\, .
\end{equation}
More precisely, we will show that the Pfaffian brane is charged under
this D3 in the Hamiltonian formalism, so we take the boundary to be of
the form $\cM^3\times\bR_t$, with the last component the time
direction along the boundary. We choose coordinates so that the
endpoint of the Pfaffian operator is at $t=0$. Now we wrap the
putative symmetry generator D3 on $\cM^3\times\RP^1$, where $\cM^3$ is
at the boundary at $t=0$. Because the $F_5$ RR flux is self-dual, the
two D3 branes that we have introduced do not commute
\cite{Gukov:1998kn,Moore:2004jv,Freed:2006ya,Freed:2006yc}:
\[
  \label{eq:Hamiltonian-Pfaffian-charge}
  \text{D3}(\cM^3\times\RP^1)\Pf(\pt) = e^{2\pi i\, \sL(\RP^1, \RP^3)} \Pf(\pt) \text{D3}(\cM^3\times\RP^1) = - \Pf(\pt) \text{D3}(\cM^3\times\RP^1)
\]
where $\Pf(\pt)$ is the D3 brane representing the Pfaffian operator,
$\sL(\RP^1,\RP^3)=\frac{1}{2}$ is the linking pairing between the
given cycles of $\RP^5$, and we have used that the given branes
intersect at a point on the $t=0$ spatial slice on
$X^5$. Equation~\eqref{eq:Hamiltonian-Pfaffian-charge} is the
Hamiltonian version of the statement that the Pfaffian operator is
charged under $\text{D3}(\cM^3\times\RP^1)$, as claimed. This
discussion can be generalised straightforwardly to show that the
branes~\eqref{eq:brane-ansatz} and \eqref{eq:NS5-brane-ansatz} do
indeed give the expected charges to the $W_V$ and $H_V$ lines of the
$SO(4N)$ theory, as claimed.

\newpage

\section{TQFT stacking from Wess-Zumino couplings}

Our task in this section will be to deduce the non-invertibility of
the symmetry generators of the theories in~\eqref{GlobalForms} from
our assumption that symmetry generators are represented
holographically by branes.

\subsection{Fluxes and twisted differential cohomology}

Our basic tool will be differential cohomology. We refer the reader to
\cite{Apruzzi:2021nmk} for a review of the basic techniques and
notation that we use. The analysis in this paper has some novelties
with respect to the discussion in \cite{Apruzzi:2021nmk}, which we now
discuss.

The main difference is that we will be working with \emph{twisted}
differential cohomology. The twisted and untwisted cohomology groups
of $\RP^5$ were given in~\eqref{RP5Coho} above. The ring structure
induced by the cup product for $\RP^5$ can be obtained by adapting the
discussion in Lemma 1 of \cite{twisted-cohomology-ring} (see also
\cite{Thom1952}). It is most easily described by adjoining the twisted
and untwisted cohomology groups
\begin{equation}
  \label{eq:twisted-ring}
  \begin{aligned}
    \mathsf{H}^*(\RP^5) & = H^*(\RP^5; \bZ)\oplus H^*(\RP^5; \tbZ) \\ & = \bZ[t_1,u_5]/(2t_1,t_1^6,u_5^2)\, .
  \end{aligned}
\end{equation}
That is, we have free components of degree 0 and 5, and $\bZ_2$
torsional components of degrees 1 to 5, generated by $t_1^n$. In
particular, taking an even number of powers of $t_1$ gives an
untwisted class, while taking an odd number of powers gives a twisted
one. In what follows we will use the notation $t_{2n+1}=t_1^{2n+1}$
for twisted classes and $u_{2n}=t_1^{2n}$ together with $u_5$ for
untwisted ones.

We denote by $\diff t_k$ a flat differential cohomology class with
characteristic class $t_k$, which we denote by $I(\diff t_k)=t_k$, and
similarly for $\diff u_k$. We note that $I(\diff t_1^2)=u_2$, and
similarly $I(\diff t_1^4)=u_4$, so perfectness of the linking pairing
on $H^2(\RP^5;\bZ)\times H^4(\RP^5;\bZ)=\bZ_2\times\bZ_2$ implies that
\begin{equation}
  \label{eq:pairing}
  \int_{\RP^5} \diff t_1^6 = \int_{\RP^5} \diff u_2\star \diff u_4 = \frac{1}{2} \mod 1\, .
\end{equation}
This equation together with the ring structure~\eqref{eq:twisted-ring}
will be our workhorses in what follows.

Finally, before moving on to the examples, we need to know how to
represent background fluxes in terms of differential cohomology. To
lighten notation, in this section we introduce $b_2\df B_2^e$ and
$c_2\df C_2^m$. Recall that the objects charged under these
backgrounds are F1 and D1 branes, respectively, so the holographic
fluxes encoding these backgrounds are $\diff H_3$ and $\diff F_3$,
which are asymptotically of the form $\diff H_3 = b_2\star \diff t_1$
and $\diff F_3 = c_2\star \diff t_1$.  By imposing this asymptotic
form we ensure that the charged lines in the field theory acquire the
right holonomies, see \cite{Garcia-Etxebarria:2019cnb,Apruzzi:2021nmk}
for analysis of similar examples. We could also include terms
proportional to $\diff t_3$ in these expansions, but they would
correspond to a change of the gauge algebra to $\mathfrak{so}(4N+1)$
(for $\diff F_3$) of $\mathfrak{usp}(4N)$ (for $\diff H_3$)
\cite{Witten:1998xy} so we will not consider these terms
further.\footnote{When doing this sort of expansion there is an
  additional subtlety involving topologically trivial differential
  characters that is discussed at length in \cite{Apruzzi:2021nmk}. It
  will not affect our considerations, so we will ignore such terms.}
Finally, a field theory 0-form symmetry background $a_1$ for
$D_2^{(0)}$ is represented by $\diff F_5=a_1\star \diff u_4$. There
are additional terms possible in the expansion for $\diff F_5$, we
will discuss these below.

\subsection{Non-invertibles in $Sc(4N)$ from D3 branes}

We start with the case of the $Sc(4N)$ theory, where following the
analysis in \cite{Kaidi:2021xfk,Bhardwaj:2022yxj}, we expect to get
the topological defect that is non-invertible from the generator of
the 0-form symmetry in the $SO(4N)$ theory. We identified this
generator above with the D3-brane wrapped on
$\cM^3\times\RP^1=\cM^3\times S^1$. The worldvolume $U(1)$ field on
the D3 brane is odd under $-1\in SL(2,\bZ)$, because it is a
trivialisation of $B_2$, which is odd. Therefore it takes values in
the twisted cohomology group $H^2(\cM^3\times \RP^1;\tbZ)$. We have
$H^*(\RP^1;\tbZ)=\{0,\bZ_2\}$.

When computing the path integral on the D3, the field strength $F_2$
on the brane will induce D1 charge due to the Wess-Zumino
term~\eqref{eq:WZ}, while the magnetic field strength $F_2^D$ will
induce F1 charge. The Wess-Zumino action written in terms of the
electric variable $F_2$ is
\begin{equation}
  S_{\text{D3},e} = 2\pi i\int_{\cM^3\times \RP^1} \diff F_5 + \diff F_3\star (\diff \cF_2) + \diff F_1 \star (\frac{1}{2}\diff \cF_2\star \diff \cF_2 + \frac{1}{24}\diff e)\, ,
\end{equation}
with $\diff \cF_2 = \diff F_2 - \diff B_2$, $\diff F_1$ a differential
cohomology uplift of $C_0$, and $\diff e$ is a differential cohomology
uplift of the Euler class of $\cM^3\times \RP^1$. The term
proportional to $\diff F_1$ will be relevant only for analysis of
anomalies in the space of coupling constants, which we do not analyse
in this note (although this is certainly an interesting direction to
explore further).

We also need to consider couplings of the form
$\diff F_3 \star \diff B_2$. As argued in \cite{Freed:2000ta}
(elaborating on results of \cite{Taylor:2000za,Alekseev:2000ch}),
these couplings are not present when measuring the actual K-theory
charges, which is what we are ultimately interested in, so we will
simply set $\diff B_2$ to 0. A more careful treatment of this issue
would be desirable, but given that inclusion of these background
fields would not change our conclusions (since they would provide
overall invertible prefactors on the brane action in any case, even if
we included them), we will postpone a more careful treatment of this
point to future work.

With these simplifications taken into account, the relevant part of the
Wess-Zumino action for the D3 becomes
\begin{equation}
  S_{\text{D3},e} = 2\pi i\int_{\cM^3\times \RP^1} \diff F_5 + \diff F_3\star \diff F_2\, .
\end{equation}
The first term is the differential cohomology avatar of the naive
guess $\int_{\cM^3\times \RP^1}C_4$ for the flux operator in the field
theory. $\diff F_5$ is even under the $-1\in SL(2,\bZ)$ action, so its
general decomposition is of the form
$\diff F_5 = N \star \diff u_5 + a_1 \star \diff u_4 + a_3 \star \diff
u_2 + N \star \diff 1$. Here $N$ are the number of units of RR 5-form
flux on the $\RP^5$, and we have used that $F_5$ is self-dual to
relate the components of degrees 5 and 0.

In terms of this decomposition we have an effective operator in
$\AdS_5$ of the form
\[
  \label{eq:D3-holonomy}
  D(\cM^3) & = \exp\left(2\pi i \int_{\cM^3\times\RP^1} \diff F_5\right) = \exp\left(2\pi i\int_{\cM^3\times \RP^5} \diff F_5 \star \diff u_4\right) \\ & = \exp\left(\pi i \int_{\cM^3} a_3\right)\, ,
\]
where in the second equality we have used Poincaré duality on $\RP^5$
to relate $\RP^1$ to $u_4$, and in the third used that the only
non-trivial pairing in $\RP^5$ appearing after the expansion of
$\diff F_5$ is~\eqref{eq:pairing}. This is the expected formula for
the operator measuring discrete electric flux for the outer
automorphism symmetry in the $SO(4N)$ theory.

The second term is the more interesting one for our purposes. As
explained above, field theory backgrounds for the symmetry $D_2^{C,m}$
are described holographically by fluxes with asymptotic form
$\diff F_3= c_2\star\diff t_1$. Similarly, we can expand
$\diff F_2 = \gamma_1\star \diff t_1$. We then have (using the
formulas for integration on products reviewed in
\cite{Apruzzi:2021nmk})
\begin{equation}
  \label{eq:D3-c2}
    2\pi i \int_{\cM^3\times \RP^1} \diff F_3\star \diff F_2 = 2\pi i \int_{\cM^3} c_2  \gamma_1 \int_{\RP^1} \diff t_1 \star \diff t_1 = \pi i \int_{\cM^3} c_2 \gamma_1
\end{equation}
where in the last step we have again used the fact that the linking
pairing is perfect, so
\begin{equation}
  \int_{\RP^1} \diff t_1 \star \diff t_1 = \frac{1}{2} \mod 1.
\end{equation}

So far we have considered the charge induced on a D3 due to the gauge
field strength $F_2$. The computation above shows that it induces an
effective coupling on $\cM^3$ to the background for $D_2^{C,m}$. By
IIB S-duality, this implies that a dual field strength
$F_2^D=\phi_1\star \diff t_1$ induces a coupling of the form
\[
  \label{eq:D3-b2}
  \pi i \int_{\cM^3} b_2 \phi_1
\]
on the effective operator on $\AdS_5$. The same result can be obtained
from the effective action presented in the magnetic variables obtained
in \cite{Kimura:1999jb}.

In elementary terms, the two couplings~\eqref{eq:D3-c2} and
\eqref{eq:D3-b2} that we have just derived can be understood as
encoding the well known facts that worldvolume flux on the D3 induces
D1 charge, and magnetic worldvolume flux F1 charge. Recall that the D1
and F1 are the charged objects in the $SO(4N)$ theory before gauging
their corresponding symmetries. After gauging, they will become the
symmetry generators for the dual magnetic symmetries in the $Sc(4N)$
theory (at least if our general philosophy of identifying branes with
symmetries is correct). So what we have just shown, is that when doing
the path integral on the D3 we will have to sum over insertions of the
symmetry generators for the 1-forms of the theory. This is certainly
suggestive that condensations
\cite{Gaiotto:2019xmp,Choi:2022zal,Roumpedakis:2022aik} are going to
enter the picture after gauging.

The precise details are nevertheless somewhat subtle. In general, when
performing the path integral the standard prescription is that we
choose whether we formulate the theory in terms of electric or
magnetic variables, and then sum over the specified variables
only. From this point of view the two couplings~\eqref{eq:D3-c2} and
\eqref{eq:D3-b2} seem somewhat at odds, and it is not clear which one
we should choose. What saves the day is that this standard
prescription has to be subtly modified whenever the cohomology groups
where the electric and magnetic fluxes live contain torsional
components. In this case, as originally pointed out by
\cite{Moore:2004jv,Freed:2006ya,Freed:2006yc}, the electric and
magnetic flux operators do not commute. As shown in
\cite{Apruzzi:2021nmk} (see also \cite{Camara:2011jg} for a different
derivation of the same result) this flux non-commutativity leads to
the existence of a discrete gauge theory when the theory is
compactified on the space with torsion. The argument, adapted to the system at hand, goes as follows.

Our initial theory is four dimensional $U(1)$ Maxwell theory on the
D3, compactified on $\cM^3\times \RP^1$. We will present a Hamiltonian
quantisation analysis, so we assume that $\cM^3=\cN^2\times \bR$, and
we identify the last component with the time direction.\footnote{A
  Lagrangian derivation will appear in \cite{GEH}.}  The spatial slice
is of the form $\cN^2\times \RP^1$. There is a non-trivial $SL(2,\bZ)$
duality bundle along the $\RP^1=S^1$ direction inherited from the
$\RP^5$ background, with holonomy $-1$, which induces a
$(F_2,F_2^D)\to (-F_2,-F_2^D)$ transformation of the worldvolume gauge
field. Therefore, just as in the IIB background itself, the
worldvolume gauge fields on the D3 are valued in twisted
cohomology. In particular $H^1(\RP^1; \tbZ)=\bZ_2$, which justifies
the statement above that there is torsion in this problem.

Consider the operators $\Phi_e(a\otimes t_1)$, $\Phi_m(b\otimes t_1)$
that measure electric and magnetic fluxes on the torsional
sector. They are associated with flat, topologically non-trivial
elements of $\Tor H^2(\cN^2\times \RP^1;\tbZ)$
\cite{Freed:2006ya,Freed:2006yc}, which in our case are all of the
form $a\otimes t_1$, where $a\in H^1(\cN^2;\bZ)$ and $t_1$ is the
generator of $H^1(\RP^1;\tbZ)$. Alternatively, using Poincaré duality,
we can view these operators as the holonomy of the twisted fluxes
$\diff F_2$ and $\diff F_2^D$ on cycles $\alpha\times \widetilde{\pt}$
and $\beta\times \widetilde{\pt}$, where $\alpha$ and $\beta$ are
Poincaré dual to $a$ and $b$ in $\cN^2$, and $\widetilde{\pt}$ the
(twisted) point in $H_0(\cN^2;\tbZ)$, which is Poincaré dual on $\RP^1$
to $t_1$. So we have
\[
  \Phi_e(a\otimes t_1) = \exp\left(2\pi i \int_{\alpha} \gamma_1 \int_{\widetilde{\pt}}\diff t_1 \right) = \exp\left(2\pi i \int_\alpha \gamma_1\int_{\RP^1} \diff t_1^2\right) = \exp\left(\pi i \int_{\alpha} \gamma_1\right)
\]
and similarly
\[
  \Phi_m(b\otimes t_1) = \exp\left(\pi i \int_\beta \phi_1\right)\, .
\]
Now, it follows from the general analysis of
\cite{Moore:2004jv,Freed:2006ya,Freed:2006yc} that
\[
  \Phi_e(a\otimes t_1) \Phi_m(b\otimes t_1) = (-1)^{\int_{\cN^2} a b}
  \Phi_m(b\otimes t_1) \Phi_e(a\otimes t_1)\, ,
\]
or equivalently, formulating everything in terms of homology on
$\cN^2$ (and abusing notation slightly):
\[
  \Phi_e(\alpha) \Phi_m(\beta) = (-1)^{\alpha\cdot \beta} \Phi_m(\beta)\Phi_e(\alpha)\, .
\]
These commutation relations are precisely those of a $\bZ_2$
theory. We can represent this theory by a gauge theory on the two
fields $\gamma_1$, $\phi_1$ with action \cite{Banks:2010zn}
\[
  S_{\bZ_2} = \pi i \int_{\cM^3} \gamma_1\delta \phi_1\, .
\]
We have identified the fields appearing in the Lagrangian with
$\gamma_1$ and $\phi_1$ since these are precisely the fields whole
holonomies are measured by the operators in the theory, by
construction.

Assembling all the pieces together, we find that the effective
partition function on the D3, seen as an 3-surface dynamical object on
$X^5$, is (up to an overall normalisation)
\[
  \label{eq:N-Sc}
  \cN(\cM^3) = D_3^{(0)}(\cM^3) \cdot \int \cD\gamma_1\cD\phi_1
  \exp\left(\pi i \int_{\cM^3} \gamma_1\delta \phi_1 + c_2\gamma_1 +
    b_2\phi_1 \right)\, .
\]
The path integral over $\gamma_1,\phi_1$ is the remnant of the $U(1)$
YM path integral in this torsional setting. This is precisely the
non-invertible operator found in \cite{Bhardwaj:2022yxj}.

\subsubsection*{Fusion rules}

Now that we have a full description of the symmetry defect, including
its TQFT sector, we can derive the fusion rules for the extended
operators in the $Sc(4N)$ theory, in particular showing that
$\cN(\cM^3)$ is a non-invertible operator of the $Sc(4N)$
theory. Since the TQFT that comes out of the brane dynamics is
identical to the one conjectured in \cite{Bhardwaj:2022yxj}, the rest
of our derivation of the fusion rules can proceed exactly as in that
paper (and the similar analysis in \cite{Kaidi:2021xfk}). We include
the details of the argument for completeness and convenience for the
reader, and then offer some comments reinterpreting some of the
features of the computation from a brane perspective.

Consider first the fusion of two copies of $\cN(\cM^3)$. Each defect
comes with its own $\bZ_2$ TQFT, so we have two sets of dynamical
fields:
\[
  \cN(\cM^3)\times \cN(\cM^3) = \int\cD\gamma_1\cD\phi_1\cD\gamma_1'\cD\phi_1' \exp\left(\pi i
    \int_{\cM^3} \gamma_1\delta \phi_1 + \gamma_1'\delta \phi_1' + c_2(\gamma_1+\gamma_1') + b_2(\phi_1+\phi_1')\right)\, .
\]
Switching to new variables $\gamma_1$,
$\hat\gamma_1\df\gamma_1+\gamma_1'$, $\phi_1$,
$\hat\phi_1\df \phi_1+\phi_1'$, the action becomes
\[
  \cN(\cM^3)\times \cN(\cM^3) = \int\cD\gamma_1\cD\phi_1\cD\hat\gamma_1\cD\hat\phi_1 \exp\left(\pi i
    \int_{\cM^3} \hat\gamma_1\delta \hat\phi_1 + \hat\gamma_1\delta \phi_1 + \gamma_1\delta \hat\phi_1 + c_2\hat\gamma_1 + b_2\hat\phi_1\right)\, .
\]
We can integrate $\phi_1$ and $\gamma_1$ out, which imposes
$\delta\hat\gamma_1=\delta\hat\phi_1=0$, so
$\hat\gamma_1\delta\hat\phi_1=0$. We then have
\[
  \cN(\cM^3)\times \cN(\cM^3) = \int\cD\hat\gamma_1\cD\hat\phi_1
  \exp\left(\pi i \int_{\cM^3} c_2\hat\gamma_1 +
    b_2\hat\phi_1\right)\, .
\]
Poincaré dualising $\hat\gamma_1$ and $\hat\phi_1$ to
$\Gamma,\Phi\in H_2(\cM^3;\bZ)$, this can be rewritten as
\[
  \label{eq:condensations}
  \cN(\cM^3)\times \cN(\cM^3) = \sum_{\Gamma,\Phi\in H_2(\cM^3;\bZ)}
  D_2^{\ssc,e}(\Gamma) D_2^{\sss,m}(\Phi)
\]
where $D_2^{\ssc,e}$ and $D_2^{\sss,m}$ are the 1-form symmetry generators
of the $Sc(4N)$ theory. (The notation is explained below.) So $\cN$ is
indeed a non-invertible defect in the $Sc(4N)$ theory, since the right
hand side is a sum of operators.

This was the derivation in \cite{Bhardwaj:2022yxj}. Holographically,
the physical meaning of the computation can be understood as
follows. We have argued that the defect $\cN(\cM^3)$ corresponds to a
D3 wrapping $\cM^3\times\RP^1$, including its quantum dynamics. The
effect of the quantum dynamics is to sum over induced charges, which
in this case means summing over D3/F1 and D3/F1 bound states. (The
precise way in which this sum happens involves, as shown above, a
$\bZ_2$ gauge theory.) If there was no sum, but only a fixed induced
charge (the trivial one, say), then taking the square would lead to a
complete annihilation of the $\bZ_2$ charges, and therefore a trivial
operator. Since there is a sum involved some of the cross-terms in the
square of the sum will lead to incomplete annihilations, leaving a sum
over F1 and D1 insertions along the worldvolume of the D3. The D3
charge is always there no matter the induced charge, and disappears,
so only the sum over D1 and F1 insertions remains.

In order to show that this physical process does indeed
produce~\eqref{eq:condensations}, all we need to verify is that the
symmetry generators of the $Sc(4N)$ theory are the F1 and D1. This is
immediate, since they are the genuine lines in the $SO(4N)$ theory,
and we are gauging the symmetry they are charged under, so they become
the magnetic symmetry generators in the gauged theory. It is also
instructive to derive it from the $\Spin(4N)$ starting point. The
\mbox{$\sss$-spinor} Wilson line $W_\sss$, given by a D5 brane wrapped on
$\RP^4$, is neutral under $\bZ_2^\sss$ (recall our conventions from
footnote~\ref{fn:conventions}), which is the symmetry that we gauge to
go to $Sc(4N)$. So the corresponding charge operator, the D1 on
$\widetilde{\pt}$, survives as a charge operator on the $Sc(4N)$
theory. We have denoted it above by $D_2^{\ssc,e}$. On the other hand
the $\ssc$-spinor and vector Wilson lines are not invariant, due to the
presence of fundamental strings in them, which are not invariant under
$\bZ_2^\sss$. So after gauging $\bZ_2^\sss$ the fundamental string on
a twisted point in $\RP^5$ becomes the second (magnetic) symmetry
generator in the $Sc(4N)$ theory, which we have denoted above by
$D_2^{\sss,m}$.

We are finally left with the task of determining the fusion of
$\cN(\cM^3)$ with the one-form symmetry generators $D_2^{\ssc,e}(\Gamma)$
and $D_2^{\sss,m}(\Phi)$. Consider for example $D_2^{\ssc,e}(\Gamma)$. We
have just argued that it corresponds to a D1 brane on
$\Gamma\times\widetilde{\pt}$. Fusing it with $\cN(\cM^3)$, which
involves a sum over induced D1 branes wrapping the Poincaré dual
$\PD[\gamma_1]\times\widetilde{\pt}$ to $\gamma_1$ amounts to shifting
$\gamma_1\to\gamma_1+\PD[\Gamma]$ in~\eqref{eq:N-Sc}. But this can
clearly be reabsorbed in a change of variables, giving back
$\cN(\cM^3)$. So
\[
  \cN(\cM^3)\times D_2^{\ssc,e}(\Gamma) = \cN(\cM^3)\, .
\]
An identical argument shows
\[
  \cN(\cM^3)\times D_2^{\sss,m}(\Phi) = \cN(\cM^3)\, .
\]

\medskip

We have shown that the $\cN(\cM^3)$ operators of the $Sc(4N)$ theory
are non-invertible, and are represented holographically by D3
branes. A small puzzle remains: our starting point was that the bulk
of the holographic dual was the same for all global forms, so the same
D3 brane appears in the bulk of all theories with the same local
dynamics, including theories that are not expected to have
non-invertible symmetries. The reason that the D3 does not lead to
non-invertible symmetries in some cases has to do with boundary
conditions (as it should, as this is the only thing that is different
in the various cases). Consider for instance the $SO(4N)$ theory,
where the D3 on $\RP^1$ is also a symmetry operator, implementing the
outer automorphism. As we push to the boundary, we obtain an operator
of the form~\eqref{eq:N-Sc}, but with a crucial difference: the IIB
$B_2$ and $C_2$ fields have a Dirichlet boundary condition in this
case, so they are not dynamical but instead they provide backgrounds
for the global 1-form symmetries for the $SO(4N)$ theory. So the term
\[
  \int\cD\gamma_1\cD\phi_1 \exp\left(\pi i
    \int_{\cM^3} \gamma_1\delta \phi_1 + C_2\gamma_1 + B_2\phi_1
  \right)
\]
in~\eqref{eq:N-Sc} (where we have capitalised $B_2$ and $C_2$ to
indicate that now they are fixed background fields) does not depend on
any dynamical field in the $SO(4N)$ theory, so it is essentially
trivial as an operator of the $SO(4N)$ theory (it can be taken out of
the path integral). In this case it is consistent to split it off from
the invertible part $D_3^{(0)}(\cM^3)$, which can meaningfully be
considered in isolation.

\subsection{4d $PO(4N)$ and $\Pin^+$ non-invertibles}

The other two theories with non-invertible symmetries
in~\eqref{GlobalForms} can be analysed in a very similar way.

Let us start with the $PO(4N)$ case. Here we gauge $D_3^{(0)}$ and
$D_2^{B,e}$, so we expect the non-invertible 2-surface operator to be
associated with $D_2^{C,m}$, which we argued above is given by a D5
brane wrapping $\RP^4\subset\RP^5$. We will need the twisted and
untwisted cohomology groups of $\RP^4$, these are
\[
  H^*(\RP^4;\bZ) & = \{\bZ, 0, \bZ_2, 0, \bZ_2\}\\
  H^*(\RP^4;\tbZ) & = \{0,\bZ_2,0,\bZ_2, \bZ\}\, .
\]
(The second line follows from analysing the twisted Gysin sequence in
\cite{Thom1952}.) As above, $\diff F_2$ is in the twisted sector, so
it expands as $\gamma_1\otimes \diff t_1$, but its magnetic dual
$\diff F_4^D$ is now \emph{untwisted}: this is needed to be able to
write a kinetic term on the twisted $\RP^4$. It therefore has an
expansion of the form
$\diff F_4^D = \phi_4\star 1 + \phi_2\star \diff u_2 + \phi_0\star
\diff u_4$.

In the electric frame the action on the D5 is of the form
\[
  S = \int_{\cM^2\times \RP^4} \diff F_7 + \diff F_2 \star \diff F_5 +
  \left(\frac{1}{2} \diff F_2^2 + \frac{1}{24}\diff e\right)\star \diff F_3
    + \ldots
\]
where the missing terms are proportional to $\diff F_1$, so we will
ignore them. The term proportional to $\diff F_2^2\star\diff F_3$ does
not contribute for degree reasons, as it goes as $\diff t_1^3$.  The
curvature term $\diff e\star \diff F_3$ could in principle contribute,
but it does not depend on the electric field so it will not enter our
considerations. We are left with the first two terms. The first one
does clearly contribute, and leads to the expected ``naive'' 2-surface
holonomy operator on $\cM^4$, entirely analogously to the discussion
around~\eqref{eq:D3-holonomy}. The second term is also
interesting. Given our expansion of $\diff F_5$ above, there is a
single non-vanishing contribution of the form
\[
  2\pi i \int_{\cM^2\times\RP^4} (\gamma_1\star \diff t_1)\star (N \star \diff u_5 + a_1 \star \diff u_4 + a_3 \star \diff
u_2 + N \star \diff 1) = \pi i \int_{\cM^2} \gamma_1 a_1\, ,
\]
where we have used that $\RP^4$ is Poincaré dual to $t_1$ in
$\RP^5$. This is the statement that worldvolume flux $F_2$ induces D3
charge. The magnetic flux $F_4^D$ will induce F1 charge (by a
generalisation of the analysis in \cite{Kimura:1999jb}), via a
coupling of the form
\[
  2\pi i \int_{\cM^2\times \RP^4} \diff F_4^D \star \diff H_3 = \pi i \int_{\cM^2} \phi_0 b_2\, ,
\]
where we have used the expansion $\diff H_3 = b_2\star \diff t_1$ as
above.

All that remains is to obtain the prescription for how to sum over
electric and magnetic fluxes. As above, flux non-commutativity can be
used to argue that there is an effective $\bZ_2$ gauge theory on
$\cM^2$ with action
\[
  i\pi \int_{\cM^2} \gamma_1 \delta \phi_0\, .
\]
The only new subtlety in this derivation comes from the fact that on
$\RP^4$, being non-orientable, the perfect torsional pairing is
between a twisted class $\diff t_1$ and an untwisted one $\diff
u_4$. An easy way to verify the existence of such a coupling is to use
Poincaré duality on $\RP^5$:
\[
  \int_{\RP^4} \diff t_1 \diff u_4 = \int_{\RP^5} \diff t_1^6 = \frac{1}{2} \mod 1\, .
\]

Putting all these terms together we obtain the topological action
\[
  \label{eq:PO}
  S_{TFT}^{PO(4N)} = i\pi \int_{\cM^2} \gamma_1\delta \phi_0 + \phi_0b_2 + \gamma_1a_1
\]
which is precisely the action proposed in \cite{Bhardwaj:2022yxj}. The
fusion algebra can be derived as above.

\medskip

Finally, in the 4d $\Pin^+(4N)$ SYM theory we gauge $D_3^{(0)}$ and
$D_2^{C,m}$, so the non-invertible surface defects are realised as
NS5-branes on $\RP^4 \times \cM^2$. The worldvolume theory on the NS5
is just as on the D5, but the gauge fields couple to the S-dual
supergravity fields. We can therefore write down the answer
immediately from~\eqref{eq:PO}:
\[
  \label{eq:Pin^+}
  S_{TFT}^{\Pin^+(4N)} = i\pi \int_{\cM^2} \gamma_1\delta \phi_0 + \phi_0c_2 + \gamma_1a_1\, .
\]

\acknowledgments

I thank Saghar Hosseini for related discussions, and Michele Del
Zotto, Ben Heidenreich and Sakura Schäfer-Nameki for initial
collaboration and discussions. This work is supported by the Simons
Foundation via the Simons Collaboration on Global Categorical
Symmetries, and by the STFC consolidated grant ST/T000708/1. I would
also like to thank the Perimeter Institute, where this work was
initiated during the 2022 workshop on Global Categorical
Symmetries. Research at Perimeter Institute is supported in part by
the Government of Canada through the Department of Innovation, Science
and Economic Development Canada and by the Province of Ontario through
the Ministry of Colleges and Universities.

\bibliography{refs}
\bibliographystyle{JHEP}

\end{document}